\documentstyle{mn}

\def\spose#1{\hbox to 0pt{#1\hss}} 
\def\simlt{\mathrel{\spose{\lower 3pt\hbox{$\mathchar"218$}}
     \raise 2.0pt\hbox{$\mathchar"13C$}}}
\def\simgt{\mathrel{\spose{\lower 3pt\hbox{$\mathchar"218$}}
     \raise 2.0pt\hbox{$\mathchar"13E$}}}

\begin{document}

\title[The X-ray off-state of CAL~83]
 {The X-ray off-state of the supersoft source CAL~83 and its interpretation}

\author[C.\ Alcock et al.]
{C. Alcock$^{1,2}$, R.A. Allsman$^{3}$, D. Alves$^{1}$, T.S. Axelrod$^{1,4}$,
D.P. Bennett$^{1,2}$,  \and P. A. Charles$^{5}$, 
K.H. Cook$^{1,2}$, K.C. Freeman$^{4}$, K. Griest$^{2,6}$, J. Guern$^{2,6}$,
\and M.J. Lehner$^{2,6}$, M. Livio$^{7}$, 
S.L. Marshall$^{2,8}$, B.A. Peterson$^{4}$, M.R. Pratt$^{2,8}$, \and 
P.J. Quinn$^{4}$, A.W. Rodgers$^{4}$, 
K.A. Southwell$^{5}$, C.W. Stubbs$^{2,8,9}$, \and 
W. Sutherland$^{5}$ and D.L. Welch$^{10}$\\
$^1$ Lawrence Livermore National Laboratory, Livermore, CA 94550, USA\\
$^2$ Center for Particle Astrophysics, University of California, Berkeley, 
CA 94720, USA\\
$^3$ Supercomputing Facility, Australian National University, Canberra, A.C.T. 
0200, Australia\\
$^4$ Mt.  Stromlo and Siding Spring Observatories, 
Australian National University, Weston, A.C.T. 2611, Australia\\
$^5$ University of Oxford, Department of Astrophysics, Nuclear \& Astrophysics
Laboratory, Keble Road Oxford, OX1 3RH\\
$^6$ Department of Physics, University of California,
San Diego, CA 92093, USA\\
$^7$ Space Telescope Science Institute, 3700 San Martin Drive, Baltimore, 
MD~21218, USA\\
$^8$ Department of Physics, University of California, Santa Barbara, CA 93106, 
USA\\
$^9$ Departments of Astronomy and Physics, University of Washington, Seattle, 
WA 98195, USA\\
$^{10}$ Department of Physics and Astronomy, Mc~Master University, Hamilton, 
Ontario, Canada, L8S~4M1\\}

\date{Received
      in original form	}

\maketitle

\begin{abstract}

We consider simultaneous optical data obtained during the recent 
X-ray turn-off of CAL\,83. Combining the optical behaviour with the observed 
X-ray decay time, we show that a model of cessation of steady nuclear burning 
is viable if the white dwarf is massive. Our model provides a 
natural explanation for the subsequent return of 
the supersoft X-ray emission. 

\end{abstract} 


\section{Introduction}

CAL~83 is considered to be a prototypical supersoft X-ray 
source (SSS; see e.\,g.\ 
Kahabka \& Tr\"umper 1996 for a review of these objects). Since its 
discovery by the {\it Einstein} X-ray Observatory 
(Long, Helfand \& Grabelsky, 1981), the source appeared to be fairly 
constant in a number of {\it ROSAT} observations (e.\,g.\ Greiner, Hasinger 
\& Kahabka 1991), but it was found to exhibit variability in some 
{\it Einstein} observations (Brown et.\ al 1994) and in the UV 
(Crampton et al.\ 1987; Bianchi \& Pakull 1988). Recently, Kahabka (1996) 
discovered the source to 
be in an unexpected X-ray off-state, three weeks after it was found to be in 
a high state. However, a later {\it ROSAT} pointing, obtained $\sim 100$~d 
after the X-ray off-state, revealed the supersoft X-ray emission to have 
returned to its high state level (Kahabka, Haberl \& Parmar 1996). 
Until now, CAL\,83 has been considered to be a persistent SSS. 
The currently accepted model for the SSS is that of steady nuclear burning on 
the surface of an accreting white dwarf (van den Heuvel et al.\ 1992; Southwell 
et al.\ 1996). 

In the present letter, we discuss several possible models for the surprising 
X-ray off-state. We then use new optical observations to distinguish 
between the models and to determine their viability. Finally, we use what we 
regard as the most plausible model to place constraints on the parameters of 
the system. 

\section{The X-ray turn-off and probable models}

Kahabka (1996) presents three observations with the {\it ROSAT} HRI. On 
1996 March~28, the count rate was $0.206\pm0.011$~s$^{-1}$, on April~7 it 
was $0.156\pm0.012$~s$^{-1}$ and on April~28 it was 
$0.0047\pm0.0024$~s$^{-1}$. If it is assumed that the decline between the 
first and third observations is exponential, then a decay time of 
$\simlt 6$~d is indicated (if the decay was linear, then the characteristic 
timescale is $\simlt 20$~d). 

If the X-ray turn-off represents an intrinsic change (and is not caused, 
for example, merely by obscuration of the source; we point out later why this 
is inconsistent with the observations), then in principle, 
two models are possible. An X-ray turn-off can be caused either by a drop in 
the bolometric luminosity, or by a decrease in the effective temperature 
(or both). In the context of nuclear burning on the surface of a white dwarf, 
such a change can occur in two different 
ways (e.~g.\ Prialnik \& Kovetz 1995; 
Krautter et al.\ 1996). These are: (i) when nuclear burning stops, the white 
dwarf envelope cools and the luminosity declines back to quiescence, and 
(ii) if the photosphere of the white dwarf, which is burning nuclear fuel at 
its surface, expands (as at the onset of a nova outburst), then the effective 
temperature 
decreases, thus shifting the emitted power from X-rays to the UV/optical 
regime. Change (i) above may be expected to occur if the accretion rate 
decreases, so that it drops below the value corresponding to steady burning. 
Change (ii) can occur if the accretion rate increases, resulting in the 
white dwarf expanding to red giant dimensions (e.~g.\ Nomoto 1982). 

In order 
to determine which of these two possibilities might have occurred, we 
consider optical observations covering the X-ray turn-off period, obtained 
via the MACHO project (e.~g.\ Alcock et al.\ 1995). The 
relevant part of the optical light curve is shown in Fig.~1, where we have 
marked the dates of the X-ray observations of Kahabka (1996) with vertical 
dotted lines. Whilst the 
optical coverage is rather sparse, the following points can be noted. The 
first X-ray observation (on March~28, at which time the X-rays were on), is 
about ten days after a time when the optical was high. The second X-ray 
observation, in which the X-ray count rate was only slightly reduced, is only 
a few days before what is a clear optical low state. The third X-ray 
observation, in which the X-rays were off, occurs when the optical has risen 
from a minimum, almost to its normal level. The general impression is 
therefore that of an optical minimum, which precedes by $\sim 10-15$~d an 
X-ray minimum. Such a behaviour is consistent with model (i) above, namely 
the following sequence of events: ({\it a}) the mass transfer rate (and 
therefore the accretion rate) decreases, ({\it b}) the optical luminosity
(which is generated mainly in the accretion disc; e.g.\ Crampton et al.\ 1996) 
decreases, and ({\it c}) 
the steady nuclear burning ceases, causing  
a cooling and contraction of the hot surface layers, and a concomitant 
turn-off in the X-rays. This behaviour is then similar to that observed in 
the decline phases of classical novae (e.~g.\ Krautter et al.\ 1996). 
Furthermore, we expect that the X-rays should turn back on, 
shortly after the optical returns to its pre-minimum level and the steady 
nuclear burning resumes. This is consistent with the most recent {\it 
ROSAT} pointing taken $\sim 100$~d after the X-ray off state, when a count 
rate of $0.21\pm0.04$~s$^{-1}$ was detected (Kahabka, Haberl \& Parmar 1996). 

It should be noted that in model ({\it ii}) above, an entirely different 
behaviour would have been expected, namely a decline in the X-rays 
which accompanies {\it an increase in the optical}; this is inconsistent 
with the 
observations. Furthermore, a scenario in which the decline in X-rays and 
optical light is caused simply by obscuration of the source is also probably 
inconsistent with the available data. This is because the supersoft X-rays 
would certainly be extinguished before any optical decline 
(e.g.\ SSS are undetectable as soft X-ray sources in the 
Galactic plane due to the high column density; van den Heuvel et al.\ 1992). 

\section{The implications of the proposed model on the system parameters}

We can attempt to use the optical and X-ray data to place some constraints 
on the system parameters. The decrease in the optical is by at least 
$0.65$~mag. Assuming that the luminosity is generated in the accretion disc, 
this represents a decrease in the mass transfer rate by a factor $\simgt 2.5$ 
(e.~g.\ Webbink et al.\ 1987). 
Given the narrowness of the steady nuclear burning strip in the 
$\dot{M}-M_{\rm WD}$ plane (Nomoto 1982), such a reduction can definitely 
result in a cessation of steady burning. 
Occasional drops in the mass transfer 
rate are often observed in nova-like variables and, in particular, in the 
group of cataclysmic variables known as VY\,Scl stars 
(e.~g.\ Honeycutt, Robertson \& Turner 1995), and in the binary 
SSS RX~J0513.9-6951 (Reinsch et al.\ 1996; Southwell et al.\ 1996). 
The VY~Scl stars are normally found in the orbital period range $3-4$~hrs, 
whereas CAL~83 has $P_{\rm orb} = 1.04$~d. Livio \& Pringle (1994) have 
suggested that the low states in the former systems may be caused by star 
spots on the surface of the secondary which cover the $L_1$ point; the short 
orbital periods are 
then expected because as the rotation rate of the star (which is coupled to 
the orbit) increases, so does the level of magnetic activity. However, this 
model is expected 
also to work for the SSS, which have longer periods, since the physical 
quantity which actually 
characterises the magnetic activity is the Rossby number, $P_{\rm
rot}/\tau_{\rm C}$, where $\tau_{\rm C}$ is the convective overturn
time in the envelope (e.g.\ Schrijver 1994). The secondary in CAL~83 is 
probably 
evolved (see van den Heuvel et al.\ 1992 for a discussion), and thus has a 
deeper convective
envelope (longer $\tau_{\rm C}$) than a main sequence star. Hence, a  
level of magnetic activity comparable to the VY~Scl stars is definitely  
possible, despite the longer periods of the SSS. 
In fact, the orbital period of CAL\,83 puts it in the range spanned by the 
magnetically active RS\,CVn stars.

Once steady burning stops, the somewhat extended white dwarf envelope cools 
and contracts. We can use the limits on the decline time for the X-rays 
($\sim 6-20$~d) to place constraints on the white dwarf mass in the system. 
A reasonable estimate of the decline time is given by $t_{3bol}$, the time 
it takes a white dwarf with nuclear burning at its surface to decline by 
three magnitudes in its bolometric luminosity. An examination of the 
results of Prialnik \& Kovetz (1995), who calculated $t_{3bol}$ for an 
extended grid, reveals that in order to obtain $t_{3bol} \simlt 20$~d, the 
white dwarf mass must satisfy $M_{\rm WD} \simgt 1.3 M_{\odot}$. In fact, a 
decline time as short as $t_{3bol} = 4.3$~d was obtained for $M_{\rm WD} = 
1.4 M_{\odot}$. 

Further confirmation of the fact that the white dwarf in the system has to be 
massive can be obtained by estimating an upper limit to the decline time, 
using the Kelvin-Helmholtz timescale of the envelope: 
\begin{equation}
\tau_{\rm{\small KH}} \approx \frac{GM_{\rm{\small WD}} \Delta m_{\rm env}}
{R_{\rm{\small WD}} L_{\rm{\small WD}}},
\end{equation}
where $L_{\rm{\small WD}}$ is the luminosity and $\Delta m_{\rm env}$ is the 
mass of the 
envelope. The latter quantity is not known, but we can obtain an upper 
limit to $\tau_{\rm{\small KH}}$ by using the envelope mass required to 
obtain a thermonuclear runaway on the surface of a {\it cold} white dwarf. The 
envelope in place during steady burning (on a hot white dwarf) can be 
significantly smaller than this (e.g.\ Prialnik \& Kovetz 1995). The envelope 
mass is given approximately (e.~g.\ Yungelson et al.\ 1995)
by:
\begin{equation}
\frac{\Delta m_{\rm env}}{M_{\odot}} \approx 2 \times 10^{-6} 
\left(\frac{M_{\rm{\small WD}}}{R_{\rm{\small WD}}^{~4}}\right)^{-0.8}. 
\end{equation}
Substituting Eqn.~2 into Eqn.~1 gives for the upper limit on the decay 
timescale (scaled with the parameter values of a $1.4 M_{\odot}$ white dwarf), 
\begin{equation}
\tau_{\rm decay} \simlt \tau_{\rm KH}  \approx 116~{\rm days} 
\left(\frac{M_{\rm{\small WD}}}{1.4 M_{\odot}} \right)^{0.2} 
\left(\frac{R_{\rm{\small WD}}}{2 \times 10^{-3} R_{\odot}} \right)^{2.2}  
\left(\frac{L_{\rm{\small WD}}}{10^{38}{\rm ~erg~s}^{-1}}\right)^{-1}. 
\end{equation} 
An examination of Eqn.~3 confirms the fact that in order to obtain the 
short observed decay time, the white dwarf needs to be very massive. 
We should note that the secondary star in CAL~83 is believed to have a mass 
of $1.5-2.0 M_{\odot}$ (e.g.\ van den Heuvel et al.\ 1992) and therefore that 
it is 
unstable to thermal timescale mass transfer even for a massive white dwarf. 

The fact that we find the white dwarf to be massive raises the
question of selection effects. An examination of the evolution of white 
dwarfs in the 
$L-T_{\rm eff}$ plane, when undergoing nuclear burning, reveals that 
white dwarfs less massive
than $\sim 0.8 M_{\odot}$ would not have been detected at all as SSS since 
they never
reach high enough effective temperatures (e.g.\ Iben 1982). 
Similarly, the luminosity of the
system is also higher the more massive the white dwarf. However, with one 
system, these selection effects should not be over-emphasised.

\section{Summary and Conclusions}

We have considered optical observations which were taken simultaneously 
with the recently discovered X-ray off-state of the supersoft X-ray source 
CAL\,83. The simultaneous observations allowed us to present a model, in 
which a drop in the mass transfer rate causes the cessation of steady 
nuclear burning, resulting in the X-ray turn-off. The steady burning resumes 
shortly after the optical returns to its pre-minimum level, causing the 
soft X-rays to turn back on. On the basis of this model, 
we predict that the white dwarf in CAL\,83 has to be very massive. 

\subsection*{Acknowledgments}

ML acknowledges support from NASA Grants NAGW~2678 and GO-05499 
at the Space Telescope Science Institute. KAS is supported by a PPARC 
studentship. We are grateful for the support given our project by the technical
staff at the Mt. Stromlo Observatory. Work performed at LLNL is 
supported by the DOE under contract W-7405-ENG. Work performed by the
Center for Particle Astrophysics personnel is supported by the NSF 
through grant AST 9120005. The work at MSSSO is supported by the Australian
Department of Industry, Science and Technology. 
KG acknowledges support from DoE OJI, Alfred P. Sloan, and Cotrell Scholar 
awards. 
CWS acknowledges the generous support of the Packard and Sloan Foundations.
WS is supported by a PPARC Advanced Fellowship.

%\clearpage
%\newpage

\begin{figure}
\caption{Optical photometry of CAL\,83 acquired from the MACHO Project 
(e.~g.\ Alcock et al.\ 1995). The relative magnitude is plotted for the 
`$B$' filter, which is approximately equivalent to the Johnson $V$ passband. 
One sigma error bars are shown. The dotted vertical lines indicate the times 
of the 3 {\it ROSAT} X-ray observations of Kahabka (1996). 
The X-rays were on for the first 
2 observations, but had switched off by the time of the third (see Sec.~2).}
\end{figure}


\clearpage
\newpage

\begin{thebibliography}{}
\bibitem[\protect\citename{}]{}
Alcock, C. et al., 1995, Phys.~Rev.~Lett., 74, 2867
\bibitem[\protect\citename{}]{}
Bianchi, L., Pakull, M.W., 1988, in A Decade of UV Astronomy with IUE, 
ESA SP-281, p.~145 
\bibitem[\protect\citename{}]{}
Brown, T., C\'ordova, F.A., Ciardullo, R., Thompson, R., 1994, ApJ, 422, 118
\bibitem[\protect\citename{}]{}
Crampton, D., Cowley, A.P., Hutchings, J.B., Schmidtke, P.C., Thompson, I.B., 
Liebert, J., 1987, ApJ, 321, 745
\bibitem[\protect\citename{}]{}
Crampton, D., Hutchings, J.B., Cowley, A.P., Schmidtke, P.C., McGrath, T.~K., 
O'Donoghue, D., Harrop-Allin, M.~K., 1996, ApJ, 456, 320
\bibitem[\protect\citename{}]{}
Greiner, J., Hasinger, G, Kahabka, P., 1991, A\&A, 246, L17
\bibitem[\protect\citename{}]{}
Honeycutt, R.~K., Robertson, J.~W., Turner, G.~W, 1995, in 
Bianchini, A.\ et al., eds, Cataclysmic Variables. Kluwer, Dordrecht, p.~75 
\bibitem[\protect\citename{}]{}
Iben, I.~Jr., 1982, ApJ, 259, 244
\bibitem[\protect\citename{}]{}
Kahabka, P., 1996, A\&A, in press
\bibitem[\protect\citename{}]{}
Kahabka, P., Tr\"umper, J.E., 1996 in 
van den Heuvel, E.P.J., van Paradijs, J., eds, Proc.\  
IAU Symp.\ 165, Compact Stars in Binaries. Kluwer, Dordrecht, p.~425
\bibitem[\protect\citename{}]{}
Kahabka, P., Haberl, F., Parmar, A.~N., 1996, IAUC 6467
\bibitem[\protect\citename{}]{}
Krautter, J., \"Ogelman, H., Starrfield, S., Wichmann, R., Pfeffermann, E., 
1996, ApJ, 456, 788 
\bibitem[\protect\citename{}]{}
Long, K.S., Helfand, D.J., Grabelsky, D.A., 1981, ApJ, 248, 925
\bibitem[\protect\citename{}]{}
Livio, M., Pringle, J.~E., 1994, ApJ, 427, 956
\bibitem[\protect\citename{}]{}
Nomoto, K., 1982, ApJ, 253, 798
\bibitem[\protect\citename{}]{}
Prialnik, D., Kovetz, A., 1995, ApJ, 445, 789
\bibitem[\protect\citename{}]{}
Reinsch, K., van~Teeseling, A., Beuermann, K, Abbott, T.M.C., 1996, A\&A, 309, 
L11  
\bibitem[\protect\citename{}]{}
Schrijver, C.~J., 1994, in Caillault, J.-P., ed, ASP Conf.\ Ser., 64, 
Cool Stars, Stellar Systems and the Sun. ASP, San Francisco, p.~328
\bibitem[\protect\uncitename{}]{}
Southwell, K.~A., Livio, M., Charles, P.~A., O'Donoghue, D., Sutherland, W., 
1996, ApJ, in press
\bibitem[\protect\citename{}]{}
van den Heuvel, E.P.J., Bhattacharya, D., Nomoto, K., 
Rappaport, S.A., 1992, A\&A, 262, 97
\bibitem[\protect\citename{}]{}
Webbink, R.F., Livio, M., Truran, J.W., Orio, M., 1987, ApJ, 314, 653
\bibitem[\protect\citename{}]{}
Yungelson, L., Livio, M., Tutukov, A., Kenyon, S.J., 1995, ApJ, 447, 656
\end{thebibliography}
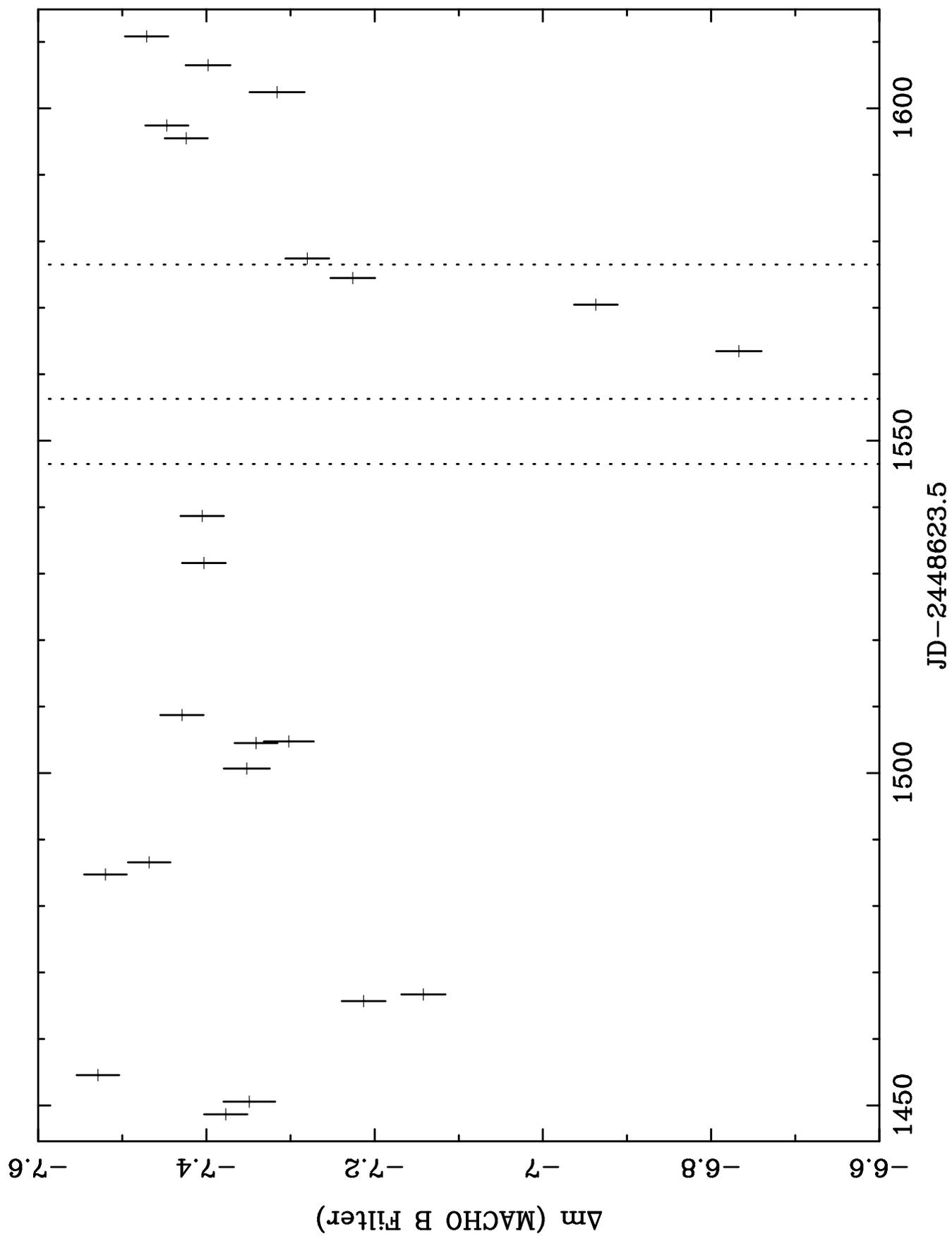
\end{document}